\documentclass[conference]{IEEEtran}

\usepackage{cite}

\ifCLASSINFOpdf
\usepackage[pdftex]{graphicx}
\graphicspath{{figures/}}
\else
\fi

\usepackage{amsmath}

\usepackage{url}

\usepackage{tabularx}
\usepackage{multirow}

\newcommand{\archname}{\emph{SegAN-CAT}}

\begin{document}
\bstctlcite{IEEEexample:BSTcontrol}
\linespread{0.98}

	\title{Brain MRI Tumor Segmentation with Adversarial Networks}	\author{
		\IEEEauthorblockN{Edoardo Giacomello}
		\IEEEauthorblockA{Dipartimento di Elettronica,\\
			Informazione e Bioinformatica\\
			Politecnico di Milano\\
			edoardo.giacomello@polimi.it}
		\and
		\IEEEauthorblockN{Daniele Loiacono}
		\IEEEauthorblockA{Dipartimento di Elettronica,\\
			Informazione e Bioinformatica\\
			Politecnico di Milano\\
			daniele.loiacono@polimi.it}
		\and
		\IEEEauthorblockN{Luca Mainardi}
		\IEEEauthorblockA{Dipartimento di Elettronica,\\
			Informazione e Bioinformatica\\
			Politecnico di Milano\\
			luca.mainardi@polimi.it}
		}

	\IEEEoverridecommandlockouts
	\IEEEpubid{\begin{minipage}{\textwidth}\ \\[12pt]
			978-1-7281-1884-0/19/\$31.00 \copyright 2019 IEEE
	\end{minipage}}
	\maketitle

	\begin{abstract}
		Deep Learning is a promising approach to either automate or simplify several tasks in the healthcare domain.
In this work, we introduce \archname, an approach to brain tumor segmentation in Magnetic Resonance Images (MRI), based on Adversarial Networks.
In particular, we extend \emph{SegAN}, successfully applied to the same task
  in a previous work, in two respects:
(i) we used a different model input and
(ii) we employed a modified loss function to train the model.
We tested our approach on two large datasets, made available by the \textit{Brain Tumor Image Segmentation Benchmark} (BraTS).
First, we trained and tested some segmentation models assuming the
  availability of all the major MRI contrast modalities, i.e., T1-weighted, T1
  weighted contrast enhanced, T2-weighted, and T2-FLAIR.
However, as these four modalities are not always all available for each patient,
  we also trained and tested four segmentation models that take as input MRIs
  acquired with a single contrast modality.
Finally, we proposed to apply transfer learning across different contrast
  modalities to improve the performance of these single-modality models.
Our results are promising and show that not only \archname\ is able to
  outperform \emph{SegAN} when all the four modalities are available, but
  also that transfer learning can actually lead to better performances
  when only a single modality is available.
 	\end{abstract}

	\IEEEpeerreviewmaketitle

	\section{Introduction}
	\label{sec:introduction}
	In the last decade,  machine learning have been applied to a large number of tasks in the healthcare domain and proved to be a promising approach
  to either automate or simplify clinical  processes that would be otherwise performed manually, requiring a great amount of time and increasing the
  possibility of human error.
Unfortunately, these applications usually rely on identifying and extracting a set of effective features from data, a very time
  consuming and problem dependent process.
In contrast, deep learning promises to solve this issue due to the capability of working directly with raw data by
  learning automatically an efficient data representation to solve the problems they are applied to~\cite{deeplearning_lecun}.
In addition, the data representation learned is often suitable also for different problems, in some cases even across different application domains.
On the other hand, these advantages come with a cost: deep learning algorithms generally require a huge amount of data to learn effective data
  representations and, hence, competent models to solve the problems they are applied to.
For this reason, several works in the literature focus on data augmentation -- i.e., how to exploit as much as possible the data available --
  and transfer learning techniques -- i.e., how to train models with data from similar problems or domains.

In this work, we focus on the MRI brain tumor segmentation task, a problem that has been widely investigated in the past few years and is the object
  of an international scientific competition, the \textit{Brain Tumor Image Segmentation Benchmark} (BraTS).
In particular, we introduce \archname, a deep learning architecture based on generative adversarial networks~\cite{goodfellow2014generative}.
We designed our architecture after \emph{SegAN}~\cite{segan}, a top performing approach in the BraTS challenge, extending it
  in two respects:
(i) we re-designed the input of one of the networks of the adversarial architecture and
(ii) we introduced an additional term in the loss function.

Our goal in this paper is threefold.
First, we aim at comparing the performance of \emph{SegAN} and \archname\ on brain tumor segmentation using as input the four MRI modalities most
  commonly acquired: T1-weighted (T1), T1-weighted contrast-enhanced (T1c),  T2-weighted (T2), and T2-FLAIR (FLAIR).
Second, as in real-world scenarios these modalities might not be all always available for each patient, we aim also at training and comparing the
  performance of four \archname\ models that use as input only a single MRI modality, i.e., one among T1, T1c, T2, and FLAIR.
Third, our aim is to investigate whether a transfer learning approach can be used to exploit the knowledge extracted from a model, previously
  trained on a \emph{source} MRI modality, to train a model that works with a different \emph{target} MRI modality.

We tested our approach by training several brain tumor segmentation models on the BraTS 2015 and BraTS 2019 datasets provided for the
  BraTS challenge.
Our results show that \archname\ is able to outperform \emph{SegAN}, especially thanks to the modified loss function employed.
As expected the results also show that using all the MRI modalities together allow to achieve a much better performance than the one achieved by
  the models that use a single MRI modality as input.
In addition, we show that the performance of single-modality models might be slightly improved by employing a transfer learning approach,
  that is by transferring the knowledge extracted from other modalities.

The remainder of this paper is organized as follows.
In Section~\ref{sec:related} we provide an overview of some of the most relevant previous works.
Then, in Section~\ref{sec:segmentation} we describe our approach and in Section \ref{sec:expdesign} we describe our experimental design.
In Section~\ref{sec:results} we report and discuss our experimental results.
Finally, we draw some conclusions in Section~\ref{sec:conclusions}.

	\section{Related Work}
	\label{sec:related}
	In this section we provide an overview of some of the previous works in the literature that applied adversarial networks to the image segmentation
  problem as well as the previous works that combined transfer learning with deep neural networks.
In addition, we also provide a brief overview of major approaches introduced in the literature to deal with missing MRI modalities.

\subsection{Adversarial Models for Segmentation}
Inspired by the recent success of \emph{Generative Adversarial Networks}~\cite{goodfellow2014generative} (GANs), Luc et al.\cite{segmentation_facebook}
  proposed a method in which a segmentation network is trained to perform pixel-wise classifications on images,
   while an adversarial network (called \textit{discriminator} or \textit{critic}) is trained to discriminate
  segmentations coming from the segmentation network and the ground truth.
This approach has been tested on the \emph{PASCAL VOC 2012}~\cite{pascal-voc-2012} and on the \emph{Stanford Background}~\cite{stanforddataset}
  datasets and the results show an improvement in segmentation performances when an adversarial loss is used.
In the medical imaging domain, multiple works applied adversarial methods to segment MRI, CT, PET and other domain specific image formats.
In particular, two previous works applied adversarial learning to brain MRI.
Moeskops et al.~\cite{segmentation_adversarial_mri} investigated the effectiveness of adversarial training and dilated convolution.
Xue et. al.~\cite{segan} proposed an Adversarial Network with a Multi-Scale loss, called \emph{SegAN}, achieving better performances compared to
  the state-of-the-art methods for brain tumor segmentation~\cite{havaei2017brain,kamnitsas2017efficient}.

\subsection{Transfer Learning}
Transfer learning investigates how to exploit a \emph{source model} trained on a \emph{source domain} for a \emph{source task} on either
  a different \textit{target domain} or a different \emph{target task}.
This usually involves either a complete or a partial retraining of the \emph{source model} to adapt it to the \emph{target domain/task}.
In particular, transferring deep neural networks across different image analysis tasks has been proved successful~\cite{transfer_neuralnetworks} and
  it seems a promising approach to deal with machine learning applications for which the amount of data is limited.

An example of the ability of CNNs to learn features that are useful for different tasks is given in~\cite{multitask_learning}, where the authors
  follow a \emph{multi-task learning} approach to demonstrate that it is possible to develop a single model to perform brain, breast and cardiac
  segmentation jointly.
When working with different MRI modalities, transfer learning has been used in~\cite{transfer_MRI} for accelerating MRI acquisition times by
  applying MRI reconstruction.
In~\cite{transfer_prostate}, the authors uses transfer learning between dataset from different institutions for the prostate gland segmentation task,
  comparing performances on different dataset sizes.
A subclass of transfer learning, called \textit{domain adaptation} studies the setting in which \textit{source} and \textit{target} data have the
  same feature space but different marginal probability functions, while the task is the same for both the domains.
For example, in~\cite{domainadaptation_ct_mri} domain adaptation is applied from CT to MRI for the lung cancer segmentation task. In~\cite{domainadaptationmri} the authors considered the issue of the different protocols adopted to acquire MRIs in the clinical practice: they used a set of MRIs as Source Domain and their follow-ups as Target Domain, studying the minimal amount of Target Domain data that is necessary to achieve acceptable performance when fine-tuning a segmentation model.

However, transfer learning applied to Adversarial Networks is still an ongoing field of research.
In~\cite{transfer_gan}, the authors focused on transfer learning using \emph{WGAN-GP}~\cite{wgangp} models and investigated
  how the selection of the network weights to transfer, the size of the target dataset, and the relation between
  source and target domain affect the final performances.
Finally, in a recent work~\cite{transfer_gan_mind2mind}, Fregier and Gouray propose a new method to perform transfer learning with WGANs~\cite{wgan}.

\subsection{Missing MRI Modalities}
A commonly used approach to address the problem of MRI segmentation in the case of missing modalities is to use \emph{image synthesis} techniques to generate artificial data.
In~\cite{generation_cGAN}, Dar et al. used \emph{Cycle GAN}~\cite{cyclegan} to generate missing modalities in a uni-modal setting,
  while Sharma and Hamarneh~\cite{generation_pix2pix} designed a multi-modal architecture where the missing modalities are considered as zero-valued inputs. In \cite{generation_flair_3d}, 3D FLAIR images are generated from T1 and later used to train a classifier together with T1 images, which led to a performance improvement.
Similarly, in~\cite{generation_simultaneous} the authors addressed the problem of missing FLAIR sequences in \emph{white matter hyper-intensity segmentation} task
  by generating the FLAIR images from T1 images while performing the segmentation at the same time.
Finally,  Ge et al.~\cite{generation_pairwise},  addressed the problem of missing modalities using  pairs of GANs: given two modalities for which there are missing samples (eg. T1 and T2), a pair of generators are trained to produce one modality from the other (eg. G\textsubscript{a}: T1 $\rightarrow$ T2, G\textsubscript{b}: T2 $\rightarrow$ T1) and the corresponding discriminators use the input of the other generator as ground truth (e.g., $ D_{a}(T2, G_{b}(T1)) $ and $ D_{b}(T1, G_{a}(T2)) $ respectively).

Another common approach to deal with missing modalities consists of training a model that is invariant to the input contrast modalities.
Working on this idea, Havaci et al.~\cite{segmentation_heteromodal} computed a latent vector for each input modality
  and then combined all the latent vectors into a single representation that accounts for every modality in input.
This method was later extended in~\cite{segmentation_permutation_invariant} to allow unlabeled inputs, i.e., to no longer specify
  which modalities are provided as input.
Finally, in~\cite{segmentation_hetermodal_vae} the authors exploited Variational AutoEncoders (VAE)~\cite{VAE}, to train an encoder for each modality and, at the same time, to generate the missing modalities.

	\section{Segmentation with Adversarial Networks}
	\label{sec:segmentation}
	Adversarial Networks became popular as \emph{generative models} (Generative Adversarial Networks), in which two networks, called \emph{generator}
  and \emph{discriminator} are alternately trained to solve a minimax game.
	In the basic formulation of GANs, the discriminator network is trained to assign higher scores to samples that comes from the ground-truth and
lower scores to sample that are generated by the generator network.
The generator is trained to produce samples that can be wrongly classified from the discriminator.
As a result, the generator learns to produce samples that follow the distribution of data of the training set from input noise vectors, that are
  drawn from a so-called \emph{latent space}.

This adversarial mechanism can be used also for solving segmentation tasks:
the generator network -- dubbed \emph{segmentator} -- is trained to produce a \emph{segmentation map} of an input image;
the discriminator takes as input both an image along with its segmentation map and is trained to distinguish generated segmentations from the
  ground truth ones.

Our approach is based on \emph{SegAN}~\cite{segan}, an adversarial network architecture, that was extended in two respects:
(i) a \emph{dice loss}~\cite{diceloss} term has been added to the multi-scale loss function employed in~\cite{segan};
(ii) we provide the discriminator with the input image \emph{concatenated} to its segmentation map, while in \emph{SegAN} the discriminator
  is provided with the input image \emph{masked} using its segmentation map.

In the remainder of this section we discuss more in details our proposed architecture (dubbed from now on \archname), the loss function used to train
  it, and the differences with respect to the original \emph{SegAN}.

\subsection{Network Architecture}
\begin{figure*}
	\begin{center}
		\includegraphics[width=0.8\textwidth]{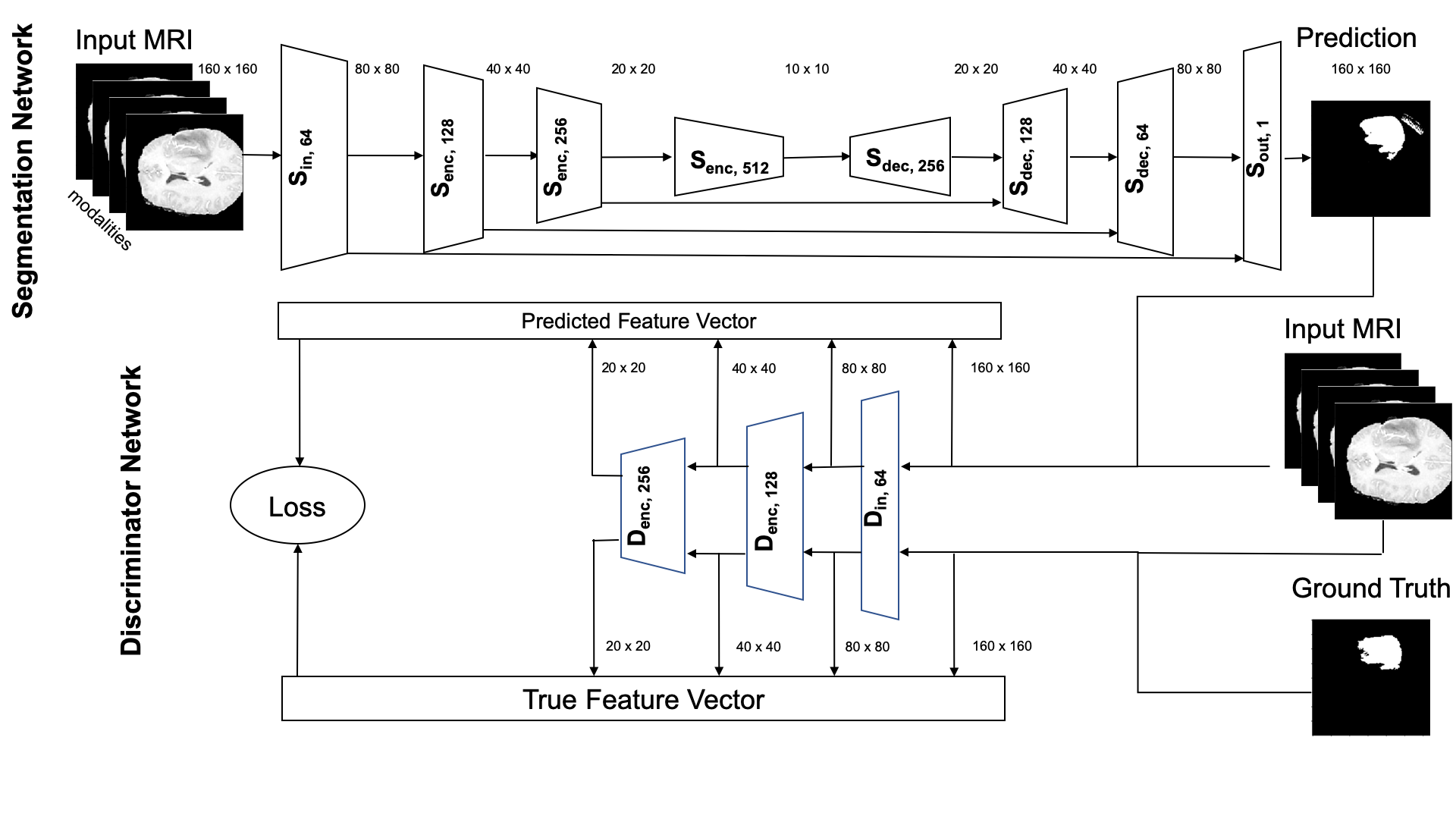}
	\end{center}
	\caption{The \archname\ architecture.}
	\label{fig:arch_overview}
\end{figure*}
The \archname, shown in Figure~\ref{fig:arch_overview}, involves a \emph{segmentation network} and a \emph{discriminator network}.
The segmentation network takes in input an MRI slice and gives in output the segmentation of the slice as a probability label map;
the slices have size 160x160x$M$, where M is the number of MRI modalities used to train the model (in this work $M$ is either 1 or 4);
the probability label maps have size 160x160, representing the probability for each pixel of the input slice of being part of the area of
 interest~\footnote{
Please notice that in this work we focused on a binary segmentation problem, but our model can be extended to multi-class segmentation problems by
  computing a probability label map of size 160x160x$L$, i.e., computing for each pixel a vector that represents the probability distribution among
  the $L$ labels.}.
Instead, the discriminator network takes in input an MRI slice and its probability label map, that can be either the one computed
   by the segmentation network or the ground-truth one:
the slice and probability label map are either combined or concatenated together as described later in this Section;
thus, the network computes a feature vector that represents a multi-scale representation of the input, which is used to compute the loss function
  during the training of the two networks.
Figure~\ref{fig:arch_overview} shows the structure of the two networks, based on the following types of computational blocks:
(i) S\textsubscript{in, k}, that is a \textit{segmentation input block} composed of a \emph{2D Convolution} layer with \textit{k} filters of size 4 and
\textit{stride=2}~\cite{lecun1990handwritten, alexnet}, followed by a \emph{LeakyReLU}~\cite{leakyrelu} activation layer;
(ii) S\textsubscript{enc, k}, that is a \textit{segmentation encoder block} has the same structure as the segmentation input block but has a \emph{batch normalization} layer~\cite{batchnorm}, before the activation layer;
(iii) S\textsubscript{dec, k}, that is a \textit{segmentation decoder block} composed of a \emph{2D Bilinear Upsampling} layer (factor=2) followed by a \emph{2D Convolution} layer with \textit{k} filters of size 3 and \textit{stride=1}, followed by a \emph{batch normalization} layer and a \emph{ReLU}~\cite{relu} activation layer;
(iv) S\textsubscript{out, k}, that is a \textit{segmentation output block} composed of a \emph{2D Bilinear Upsampling} layer (factor=2) followed by a \emph{2D Convolution} layer with \textit{k} filters of size 3 and \textit{stride=1}, followed by a \emph{Sigmoid}~\cite{mitchell} activation layer;
(v) D\textsubscript{in, k}, that is a \textit{discriminator input block} composed of a \emph{2D Convolution} layer with \textit{k} filters of size 4 and \textit{stride=2}, followed by a \emph{LeakyReLU} activation layer.
(vi) D\textsubscript{enc, k}, that is a \textit{discriminator encoder block} has the same structure as the discriminator input block but has a batch normalization layer, before the activation layer.
The convolutions weights in all \textit{discriminator blocks} are constrained between \textit{[-0.05;0.05]} for stabilizing the training process \cite{wgan}. The output of the discriminator, indicated as \emph{Feature Vector} in Figure~\ref{fig:arch_overview}, is computed by concatenating the discriminator input and the flattened output of every discriminator block.
The slope for the \emph{LeakyReLU} is 0.3; batch normalization parameters are  $\epsilon=1*10^-5$ and momentum=$0.1$ for both networks.

\subsection{Discriminator Network Input}
\label{ssec:inputs}
In the \emph{SegAN} architecture the discriminator network is given in input a \emph{masked} slice image, computed by pixel-wise multiplication of
  the label probability map and each channel of the MRI slice.
Instead, in \archname\ architecture the label probability map and each channel of the MRI slice are simply concatenated and provided to the discriminator network as input.
While the input definition used in the original \emph{SegAN} architecture is more compact, with the input definition we propose it is possible
  for the discriminator network to extract features that describe also the area of the input MRI that is \emph{not} included into the
	segmentation.
As a result, we expect \archname\ to have slightly better generalization capabilities than the architecture introduced in~\cite{segan}.

\subsection{Loss Function}
To train our networks, we use the \textit{Multiscale Adversarial Loss}, as in~\cite{segan}.
This particular loss function, applied also in generative problems~\cite{pix2pixhd}, allows to perform \textit{feature matching} between the
  ground truth and the output of segmentation network, optimizing also the network weights on features extracted at multiple
	resolutions rather than focusing just on the pixel level.
Thus, the \emph{Multiscale Adversarial Loss} is defined as follows:

\begin{equation}
\ell_\mathrm{mae}(f_D({x}), f_D({x}^{\prime})) = \frac{1}{L}\sum_{i=1}^L ||f_D^i({x}) - f_D^i({x}^{\prime})||_{1} \enspace \label{eq:multiscale_loss}
\end{equation}

where $L$ is the number of layers in the discriminator network;
$f_D^i({\cdot})$ is the output of the activation layer after the discriminator block
  at position $i+1$, e.g., $f_D^1({\cdot})$ is the discriminator input, $f_D^2({\cdot})$ is the output of the first activation layer, etc;
$x$ denotes the input of the discriminator when the label probability map is computed by the segmentation network,
  while ${x}^{\prime}$ is the input of the discriminator when the ground-truth is used.

In addition to the \emph{Multiscale Adversarial Loss}, we introduced an additional term to the loss function defined as:
\begin{equation}
	\begin{split}
	\ell_\mathrm{dice}(g, p) = 1 - \frac{2\sum_{i}^{N}p_ig_i}{\sum_{i}^{N}p_i^2+\sum_{i}^{N}g_i^2} \enspace
	\label{eq:dice_loss}
	\end{split}
\end{equation}
where $g$ is a ground truth image, $p$ denotes the predicted values and the sums run over the pixels of the image;
the definition of $\ell_\mathrm{dice}$ is based on a differentiable form of the well known \emph{S{\o}rensen-Dice} coefficient~\cite{diceloss},
  or \emph{Dice Score}, it ranges between 0 and 1, and accounts for the overlap between the segmented areas in the ground-truth and in the output of the
	segmentation network (where 0 means perfect overlap and 1 means no overlap).
The use of this term, which allows to assess the quality of the generated segmentation maps, should result in a more stable training process and eventually
in a model with better performances.

Overall, the complete objective function used to train the networks is defined as:
\begin{equation}
\begin{split}
\min\limits_{\theta_S}  &\max\limits_{\theta_D}\frac{1}{N}\sum_{n=1}^N
\ell_\mathrm{mae}(f_D(x_n \circ S(x_n)), f_D(x_n \circ y_n)) \\
 & + \ell_\mathrm{dice}(y_n, S(x_n)) \enspace
\label{eq:segan_io_loss}
\end{split}
\end{equation}
where $x_n$ is an input MRI, $y_n$ is the ground truth, and the $\circ$ operator is either a concatenation on the channel axis in \archname\
  or a pixel-wise multiplication in \emph{SegAN} (as illustrated in Section~\ref{ssec:inputs}).

	\section{Experimental Design}
	\label{sec:expdesign}
	In this work, we compared the \archname\ and the \emph{SegAN} architecture on the brain tumor image segmentation problem based on the
  \emph{Multi-modal Brain Tumor Image Segmentation Benchmark} (BraTS)~\cite{brats2015dataset}, a scientific competition organized in conjunction
  with the international conference on Medical Image Computing and Computer Assisted Interventions (\mbox{MICCAI}) since 2012.

\medskip\noindent
\textbf{Dataset.}
We used both the BraTS 2015 \cite{brats2015dataset} and the BraTS 2019 \cite{brats2019dataset} datasets.
  The Brain tumor segmentation (BraTS)  challenge focus on the evaluation of methods for segmenting brain tumors, in particular Gliomas, which are the most common primary brain malignancies.
Both the datasets we use are composed of MRI volumes of resolution 240x240x155 and four contrast modalities: T1, T1c, T2, FLAIR.
Each voxel is annotated with one among five possible labels representing the tumor sub-regions: \emph{Necrosis} (NCR), \emph{Edema} (ED),
  \emph{Non-Enhancing Tumor} (NET), \emph{Enhancing/Active Tumor} (AT) and \emph{Everything Else}.
The BraTS 2015 dataset consists of 220 high grade gliomas (HGG) and 54 low grade gliomas (LGG) MRIs.
The BraTS 2019 is the latest revision of the BraTS dataset and it consists of a total of 335 MRI volumes (259 HGG, 76 LGG);
The major similarities and differences between the two datasets are:
(i) both datasets contain 30 human-annotated MRIs from a previous dataset, BraTS 2013;
(ii) the remaining volumes in BraTS 2015 are a mixture of pre and post-operative scans that have been labeled by an ensemble of algorithms and
  later evaluated by a team of human experts, while in BraTS 2019 all the post-operative scans have been discarded and all the volumes have been
  manually re-labeled;
(iii) additional data from different institutions have been included in BraTS 2019;
(iv) in BraTS 2019, the \emph{Non-Enhancing Tumor} label has been merged with \emph{Necrosis} due to a bias that exists in the evaluation of that area.

\medskip\noindent
\textbf{Data Preprocessing.}
Since images in both datasets have an isotropic resolution of 1mm\textsuperscript{3} per voxels we do not perform any further spatial processing to
  data.
As done in~\cite{segan}, we center-crop each MRI to a 180 x 180 x 128 volume in order to remove black regions while keeping all the relevant data.
For each MRI volume, we clip voxel values to the 2\textsuperscript{nd} and 98\textsuperscript{nd} percentile in order to remove outliers,
  then we apply Feature Scaling~\cite{featurescaling} to normalize the intensity range between 0 and 1.
Finally, we split the data in Training/Validation sets (respectively of size 80\% and 20\%) using stratified sampling to keep balanced HG and LG
  subjects within each subset.
In Brats2019 dataset we also apply stratified sampling based on the institution that provided the data.

\medskip\noindent
\textbf{Task Definition.}
BraTS2015 challenge defines 3 regions of the tumor based on the combination of ground truth labels.
Instead, in our experimental analysis, we focused only on one region, that roughly corresponds to the \emph{Whole Tumor} region in BraTS challenge.
Accordingly, we re-labeled as $1$ all the voxels labeled as NCR, ED, NET, or AT in BraTS  datasets, and as $0$ the remaining ones.

\medskip\noindent
\textbf{Evaluation.} For assessing the performances of our models, we define 3 metrics derived from the well-known \emph{Confusion Matrix} \cite{errormatrix}.
  To evaluate the number of True Positives (TP), False Positives (FP), True Negatives (TN) and False Negatives (FN) and thus to compute the
    metrics, we threshold (using 0.5 as threshold value) the output of the segmentation network in order to obtain a binary classification for each pixel.
Despite our models work on a single MRI slice at a time, we compute the metrics by considering the TP, FP, TN, and FN on every slice of the same MRI volume.
In particular we considered the following three metrics:
(i) the \emph{precision} ($TP/(TP + FP)$), which measures how many voxels, classified as a tumoral lesion, are effectively part of the tumor;
(ii) the \emph{sensitivity} ($TP/(TP + FN)$), which measures how many voxels, that are part of the tumor, are correctly identified as such;
(iii) the \emph{dice score} ($2TP/(2TP  + FP + FN)$), which measures the overlap between the pixels of a binary ground truth and a given prediction.

In our experiment we selected our best model according to the dice score value as it allows to account for both False Positives and False Negatives,
  leading to a better assessment of the segmentation quality with respect to the precision and sensitivity scores.

\medskip\noindent
\textbf{Transfer Learning.}
In our experimental analysis, we used the most straightforward approach to transfer a \archname\ model trained from MRIs that include
  a single modality (\emph{source}) to train a model from MRIs that include a single but \emph{different} modality (\emph{target}):
we took the segmentation and discriminator networks trained on the source modality and re-trained (or \emph{fine-tuned}) them on the
  target modality.
In particular, we studied two different fine-tuning process to retrain the networks on the target modality:
(i) we adapted all the weights of both the segmentation and the discriminator networks;
(ii) we adapted all the weights of the segmentation network and only the weights of the input layer of the discriminator network.

Although keeping the several layers of discriminator fixed during the fine-tuning might prevent it from adapting entirely to the target domain, we believe
  that this solution could help to retain more knowledge from the source domain while adapting the segmentation network.
This choice is motivated by the fact that the discriminator has been found to be the most important part of an adversarial network to transfer to the
  target domain~\cite{transfer_gan}.

\medskip\noindent
\textbf{Experimental Equipment.}
We developed and trained all the models described in this paper using the Tensorflow 2.0a Docker environment~\cite{tensorflow2015-whitepaper, docker, docker_research}.
The experiments have been carried out on a server equipped with a 12GB nVidia Titan V, Intel Xeon E5-2609 CPU and 64GB of RAM.

	\section{Experimental Results}
	\label{sec:results}
	We trained both multi-modal models, i.e., models designed to work with MRIs that include different contrast modalities, as well as uni-modal models,
  i.e., designed to work with MRIs that include only one specific contrast modality.

In the first experiment, we trained three multi-modal models that use all the four contrast modalities, i.e., T1, T1c, T2, FLAIR, available in the
  datasets:
(i) a model that implements the original \emph{SegAN} architecture,
(ii) a model that implements the \emph{SegAN} architecture with the additional dice loss term, and
(iii) a model that implements completely the \archname\ architecture, i.e., with the additional dice loss term and
  the input discriminator concatenation.

Each model is trained and tested both on BraTS 2015 and BraTS 2019 datasets using a batch size of 64 slices. During the training phase, we perform data augmentation by applying \emph{random cropping}~\cite{alexnet} using a window of size 160 x 160, as proposed in \cite{segan}. During the validation phase we apply \emph{center cropping}~\cite{alexnet} to match the input size of the network, discarding most of the black border of the MRI.
All the models are trained using the same initialization seed,  RMSprop~\cite{rmsprop_coursera} (\textit{lr: 2*10\textsuperscript{-5}}), and
\emph{Early Stopping}~\cite{earlystopping} (\textit{patience = 300 epochs}) on Dice Score evaluation metric. An epoch consist of a full iteration over the dataset, i.e. approximately 28000 slices for the Brats 2015 dataset and 34000 for the Brats 2019 dataset.

\begin{table*}
  \caption{Performance of \emph{SegAN}, \emph{SegAN} with dice loss, and \archname.
    The results reported are the average of the dice score, the precision, and the sensitivity computed on each one of the MRI volume included
      in test set of BraTS 2015 and BraTS 2019.
    We reported in bold the best performance for each dataset.}
	\label{tab:scores_base_all}
	\begin{center}
		\begin{tabular}{| c | c | c | c | c | c | c  |}
			\hline
			\multirow{2}{*}{Model}  & \multicolumn{3}{c |}{BraTS 2015} & \multicolumn{3}{c |}{BraTS 2019}  \\
			\cline{2-7}
			& Dice Score & Precision & Sensitivity & Dice Score & Precision & Sensitivity\\
			\hline
			\emph{SegAN}              & 0.705          & 0.759 & 0.694  & 0.766          & 0.745 & 0.834 \\ \hline
      \emph{SegAN} w/ Dice Loss & 0.825          & 0.901 & 0.785  & 0.814          & 0.850 & 0.810 \\ \hline
			\archname                 & \textbf{0.859} & 0.882 & 0.852  & \textbf{0.825} & 0.842 & 0.835 \\
			\hline
		\end{tabular}
	\end{center}
\end{table*}
Table~\ref{tab:scores_base_all} compares the performance of all the multi-modal models trained on two datasets.
The results show that both the dice loss term and the discriminator input concatenation, introduced in the \archname, led to better performances
  on both datasets.
Results also suggest that BraTS 2019 is slightly more challenging than BraTS 2015, leading to a lower performance of all the the three models.

In the second experiment, we wanted to investigate how the information content of each contrast modality affects the model performance.
To this purpose, we trained four uni-modal models, each one using only a single contrast modality.

\begin{table*}
	\caption{Performance of \archname\ models trained from MRIs with only one contrast modality.
The results are reported for each modality as the average of the dice score, the precision, and the sensitivity computed on each one of the MRI
  volume included in test set of BraTS 2015 and BraTS 2019.
As a reference, the performance of \archname\ trained with all the contrast modalities is reported in the last row of the table.}
	\label{tab:scores_uni-modal}
	\begin{center}
		\begin{tabular}{| c | c | c | c | c | c | c  |}
			\hline
			\multirow{2}{*}{Modality}  & \multicolumn{3}{c |}{BraTS 2015} & \multicolumn{3}{c |}{BraTS 2019}  \\
			\cline{2-7}
			& Dice Score & Precision & Sensitivity & Dice Score & Precision & Sensitivity \\
			\hline
			T1     & 0.538 & 0.570 & 0.557 & 0.542 & 0.586 & 0.609 \\ \hline
			T1c    & 0.578 & 0.613 & 0.581 & 0.587 & 0.678 & 0.577 \\ \hline
			T2     & 0.721 & 0.773 & 0.724 & 0.675 & 0.828 & 0.607 \\ \hline
			FLAIR  & 0.810 & 0.858 & 0.787 & 0.763 & 0.757 & 0.810 \\ \hline
      ALL    & 0.859 & 0.882 & 0.852 & 0.825 & 0.842 & 0.835 \\
			\hline
		\end{tabular}
	\end{center}
\end{table*}
Table~\ref{tab:scores_uni-modal} compares the performance of these four \archname\ models trained using a single contrast modality.
The results show that none of the model trained with a single modality is able to achieve the same performance achieved by \archname\, which uses
  all the four contrast modalities together.
This suggests that none of the four modalities alone contains all the relevant information to identify the tumor.
However, the model trained using only the FLAIR modality is able to achieve a much better performance than all the other models on both the
  datasets.
These results are not surprising as FLAIR modality allows to identify more clearly the edema and the lesions in specific areas of the brain, due to
  the suppression of the cerebrospinal fluid in the images.

In the final experiment, we investigated whether it is possible to transfer a model trained on a specific contrast modality to a different contrast
  modality.
Thus, as preliminary analysis, we applied all the uni-modal models previously trained on each of the contrast modality to images taken with different modalities to
  evaluate the similarities among the models and to have an insight on how easily the knowledge learned from a modality could be transferred to the the others.

\begin{figure}
	\includegraphics[width=\columnwidth]{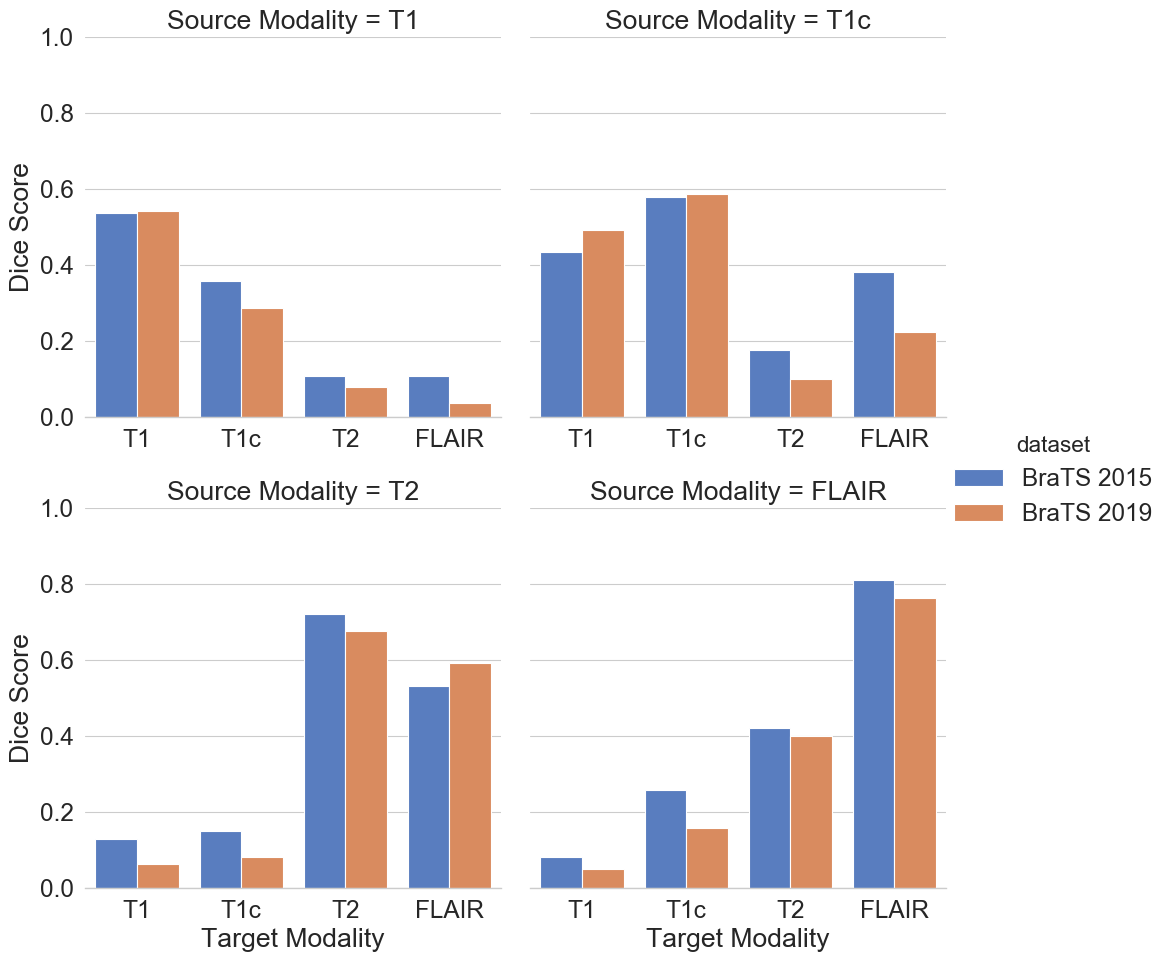}
\caption{Performance of each uni-modal model when applied to images acquired using other modalities.
The performances on both BraTS 2015 and BraTS 2019 are reported.
In addition, we also report as a reference the performance of the uni-modal models on images with the same modality the models are trained for.}
 	\label{fig:cross_dataset_performances}
 \end{figure}
Figure~\ref{fig:cross_dataset_performances} shows the performance of all the uni-modal models when applied to each modality alone.
As expected, all the uni-modal models reached the best performance on images acquired with the same contrast modality they have been trained for.
Therefore, to transfer successfully a model across the contrast modalities, it is necessary to adapt the trained networks
  with data from the target domain, i.e., the target contrast modality.
The data also show how models could be easily transferred across modalities that are known to be similar among them:
models trained on T1 and T1c perform poorly on T2 and FLAIR, while they perform much better when applied to the other modality similar
  to the one they are trained on (i.e., T1 or T1c);
the same behavior, the other way around, is found when looking at the performance of the models trained on T2 and on FLAIR.

Based on the results discussed above, we applied the transfer learning to train three uni-modal models respectively on T1, T1c, and T2 images
  by adapting a model trained on FLAIR images.
In fact, our results show that the FLAIR images account for a large part of the model performance and our aim is to understand whether the transfer
  learning process might improve the final performances.
In this experiment, the fine-tuning of the model on the target modality was limited to 300 epochs.

\begin{table*}
	\caption{Performance of the models trained by transferring the model trained on FLAIR images.
  The column \texttt{Fine Tuning} reports whether both the model networks (\texttt{S,D}) or only the segmentator and the discriminator input layer
    (\texttt{S,D\textsubscript{in}}) have been trained on the target modality.
  The results are reported for each target modality as the average of the dice score, the precision, and the sensitivity computed on each one of the MRI
      volume included in test set of BraTS 2015 and BraTS 2019.
  We reported in bold the scores that are better than the corresponding ones in Table~\ref{tab:scores_uni-modal}.}
	\label{tab:scores_transfer}
	\begin{center}
		\begin{tabular}{| c | c | c | c | c | c | c | c |}
			\hline
			\multirow{2}{*}{Target} & \multirow{2}{*}{Fine Tuning}  & \multicolumn{3}{c |}{BraTS 2015} & \multicolumn{3}{c |}{BraTS 2019}  \\
			\cline{3-8}
			 & & Dice Score & Precision & Sensitivity & Dice Score & Precision & Sensitivity  \\
			\hline
			T1   & S,D                   & 0.496           & 0.503 & 0.553 & 0.502          & 0.571 & 0.582 \\ \hline
			T1   & S,D\textsubscript{in} & \textbf{0.561}  & 0.614 & 0.576 & 0.528          & 0.558 & 0.538 \\ \hline
			T1c  & S,D                   & 0.467	         & 0.464 & 0.538 & 0.468          & 0.527 & 0.494 \\ \hline
			T1c  & S,D\textsubscript{in} & 0.577           & 0.661 & 0.541 & \textbf{0.598} & 0.705 & 0.563 \\ \hline
			T2   & S,D                   & 0.692           & 0.776 & 0.668 & 0.674          & 0.643 & 0.775 \\ \hline
			T2   & S,D\textsubscript{in} & \textbf{0.781}  & 0.818 & 0.771 & \textbf{0.741} & 0.878 & 0.681 \\ \hline
		\end{tabular}
	\end{center}
\end{table*}
Table~\ref{tab:scores_transfer} show the performance of the models trained on T1, T1c, and T2 by adapting a model trained on FLAIR.
The results suggest that it is more convenient to adapt, i.e., to train on images with the target contrast modality, \emph{only}
  the input layer of the discriminator network of the model, keeping the other layers of the discriminator network trained on images with
  the FLAIR contrast modality.
We believe that this result is due to the high instability of the adversarial learning mechanism that makes it very
  difficult to incrementally adapt a previously trained model~\cite{transfer_gan}.
Accordingly, to exploit the knowledge of the model trained on images with FLAIR contrast, we adapted on the target modalities only the
  segmentator network and the first layer of the discriminator network.
Overall, this solution often led to models that perform better or similarly to the models trained from scratch on the
  target modalities (see Table~\ref{tab:scores_uni-modal}), despite being trained for 300 epochs only.
As expected, the major benefits of transfer learning mechanism is achieved on the target modality more similar to FLAIR, i.e., T2.

  \section{Conclusions}
  \label{sec:conclusions}
  In this work, we introduced \archname, an adversarial network architecture based on \emph{SegAN}~\cite{segan}.
Our approach differs from \emph{SegAN} mainly in two respects:
(i) the loss function has been extended with a dice loss term and
(ii) the input of the discriminator network consists of a concatenation of the MRI images and their segmentation.

We applied \archname\ to an MRI brain tumor segmentation problem, the same task the \emph{SegAN} architecture was successfully applied to.
In particular, we defined a binary segmentation problem on two datasets, BraTS 2015 and BraTS 2019, used in a scientific competition organized in
  conjunction with the international conference on Medical Image Computing and Computer Assisted Interventions (MICCAI) since 2012.
The aim of this work was (i) to compare the performance of \archname\ and \emph{SegAN}, (ii) to assess the performance of uni-modal models for
  each contrast modality (i.e., T1, T1c, T2, and FLAIR), and (iii) to study the problem of transferring a previously trained model across
  different modalities.
To this purpose, we first trained \emph{SegAN}, \emph{SegAN} with a dice loss term, and \archname\ on MRIs acquired with all the four contrast modalities.
Our results on both BraTS 2015 and BraTS 2019 datasets showed that both the dice loss term and the discriminator input concatenation proposed in
  this paper allow to improve the final performance of the model.
Then, we trained one uni-modal \archname\ model for each one of the four contrast modalities in order to assess the performance that can be
  achieved using only the information available in each modality alone.
As expected, none of these four models reached the performance of the multi-modal one.
However, the model trained on FLAIR contrast modality outperformed all the others and was able to reach a performance
  quite close to the one achieved by the multi-modal model.
Thus, we investigated the possibility of transferring a model trained with FLAIR to train better model from images acquired with different contrast
  modalities.
Despite confirming that is rather difficult to use transfer learning with adversarial networks as discussed in~\cite{transfer_gan,transfer_gan_mind2mind},
  our results suggest that it is possible to successfully transfer a model trained on FLAIR contrast modality also to other modalities.
Indeed, our results show that transfer learning often allows to train uni-modal models with a better performance than the same models
  trained entirely on images with their specific contrast modality.

We believe that the transfer learning with adversarial networks is a topic worth to be further investigated in future works.
In fact, the limited amount of data is one of the major issue when it comes to the application of deep learning to healthcare domain.
So far, transfer learning has been widely used in several domains to deal with the limited availability of data, but it has been not yet fully
  exploited in the healthcare domain.
In particular, applying transfer learning to train adversarial networks is a challenging and yet rather unexplored research area that deserve
  additional investigation.
In future works we plan to investigate additional strategies to adapt a previously trained model to a target domain, such as adapting only some
  layers of the networks, alternating previously trained and untrained networks during the adapting process, and combining together networks
  trained using different contrast modalities.

	\bibliographystyle{ieeetran}

\end{document}